\documentclass[aip,amsmath,amssymb,reprint,groupedaddress]{revtex4-1}

\usepackage{graphicx}% Include figure files
\usepackage{dcolumn}% Align table columns on decimal point
\usepackage{bm}% bold math
\usepackage[utf8]{inputenc}
\usepackage[T1]{fontenc}
\usepackage{mathptmx}
\usepackage{xcolor}

\begin{document}

%\preprint{AIP/123-QED}
%\preprint{}

\title[Exciton-Plasmon Coupling ...]{Controlled Exciton-Plasmon Coupling in a Mixture of Ultrathin Periodically Aligned Single$-$Wall Carbon Nanotube Arrays}


\author{C.~M.~Adhikari~~}
\author{~I.~V.~Bondarev$\,$}\email{\vspace{-0.75cm}ibondarev@nccu.edu (Corresponding Author)}
\affiliation{Department of Mathematics \& Physics, North Carolina Central University, Durham, NC 27707, USA}


%\date{\today}

\begin{abstract}
We study theoretically the in-plane electromagnetic response and the exciton-plasmon interactions for an experimentally feasible carbon nanotube (CN) film systems composed of parallel aligned periodic semiconducting CN arrays embedded in an ultrathin finite-thickness dielectric. For homogeneous single-CN films, the intertube coupling and thermal broadening bring the exciton and interband plasmon resonances closer together. They can even overlap due to the inhomogeneous broadening for films composed of array mixtures with a slight CN diameter distribution. In such systems the real part of the response function is negative for a broad range of energies (negative refraction band), and the CN film behaves as a hyperbolic metamaterial. We also show that for a properly fabricated two-component CN film, by varying the relative weights of the two constituent CN array components one can tune the optical absorption profile to make the film transmit or absorb light in the neighborhood of an exciton absorption resonance on-demand.
\end{abstract}


\maketitle

\section{\label{sec:1} Introduction}
Carbon Nanotubes (CNs) and ultrathin films made of periodic CN arrays, provide a variety of useful physical properties that are essential for optoelectronic device applications~\cite{Bondarev2009, MaYa2011,VoEtAl2013,AaFrPe2008,WuEtAl2017}. Periodically aligned CN array films, in particular, provide stability and precise tunability of their characteristics by means of the CN diameter, intertube distance and film thickness variation, which is why they are getting more and more attention of experimental communities~\cite{RoEtal2019,HERTEL2013, ZhEtal2020,FaEtAl2017, HoEtAl2018, KuCh2017}. Self-assembled quasi-periodic finite-thickness single-wall CN (SWCN) films have been recently shown experimentally to exhibit extraordinary optoplasmonic properties such as a tunable negative dielectric response in a broad range of the photon excitation energies~\cite{RoEtal2019}. Using the Maxwell-Garnett (MG) mixing method~\cite{Ma2016} and the many-particle Green’s function formalism~\cite{Mahan2000}, we have lately explained this theoretically to be the case due to the inhomogeneous broadening effect in films composed of mixed SWCN arrays with a narrow nanotube diameter distribution~\cite{AdBo_MA2020}.

In this contribution, we focus on the exciton-plasmon coupling in a two-component mixture of periodic SWCN arrays. We start with the analysis of the dielectric response of a homogeneous quasi-2D array of parallel aligned, uniformly spaced, identical SWCNs immersed in a finite-thickness dielectric medium of static relative dielectric permittivity $\epsilon$, sandwiched between a substrate and a superstrate of relative permittivities $\epsilon_1$ and $\epsilon_2$. A schematic of the array system under study and notations of relevance are presented in the inset of Fig.~\ref{fig1}. Next, we present a numerical study on the role of the exciton-plasmon coupling in the optical absorption spectrum of a two-component mixture of the homogeneous SWCN arrays. We discuss how this coupling and thereby the absorption can be controlled by varying the relative weights of the individual components in the mixture.

The SWCNs are aligned along the $y$-axis in a dielectric layer of thickness $d$  with the intertube center-to-center distance $\Delta$. The electron charge density is distributed uniformly all over the periodic cylindrical nanotube surfaces, whereby the pairwise electron Coulomb interaction in the system of CNs can be approximated by that of two uniformly charged rings of radius $R$ of the respective $n$-th and $\ell$-th tubules~\cite{Bo2019ME}. We consider that the dielectric medium embedding a periodic array has much greater dielectric permittivity than the substrate and superstrate permittivities. In this case~\cite{Ke1980}, the Coulomb interaction in the film increases strongly as the film thickness decreases, making the Coulomb interaction independent of the vertical coordinate component~\cite{Ke1980, BoSh2017, BoMoSh2018, Bo2019ME, BoMoSh2020}. Although this vertical confinement leads to the reduction of the effective dimensionality from three to two, the thickness $d$ is still a variable parameter there to represent the finite vertical size of the system, which the dielectric response of the film depends on.

We proceed with our studies as follows. First, we determine the conductivity of an individual SWCN as a function of photon frequency using the so-called $(\bm{k\cdot p})$-band structure method~\cite{Ando2005}. Then, taking advantage of the vertical electron confinement nature of stronger Keldysh-Rytova interaction potential~\cite{Ke1980, BoSh2017, BoMoSh2018, Bo2019ME}, we determine the effective  interaction  of an ultrathin finite-thickness periodically aligned CN array and derive an expression for the dynamical dielectric response tensor of the CN-array using the low-energy plasmonic response calculation technique~\cite{Bo2019ME} and the many-particle Green's function formalism~\cite{Mahan2000}. We then mix two non-identical CN arrays using the Maxwell-Garnett mixing scheme~\cite{Ma2016} and examine what happens on a CN film's absorption
spectrum if the film has only a few non-identical homogeneous CN arrays with possible exciton-plasmon coupling presence.

Previously, the exciton-plasmon coupling was studied for a variety of nanostructured systems such as organic semiconductors, SiC nanocrystals, gold nanoshells and nanorods, metallic dimers, and hybrid metal-semiconductor nanostructures~\cite{BeEtal2004, ScEtal2013, MaEtal2011,FeEtal2007, Dai_2012, VaEtal2008, WuEtal2007, As2018}. The exciton-plasmon coupling in an individual CN has been investigated theoretically in Refs.~\cite{BoWoPo2011, BoTaWo2010, BoTaWo2009, BoWoTa2009, BoAn2012, Bon2012}, including the case with a perpendicular electrostatic field applied~\cite{BoWoTa2009, BoAn2012, Bon2012}. The interactions of excitonic states with interband plasma modes of a semiconducting SWCN were shown to result in a strong exciton-plasmon coupling that splits the exciton absorption lineshape (Rabi splitting)~\cite{BoWoTa2009}. In recent studies, such a line-splitting in CN film spectra has been observed experimentally~\cite{FaEtAl2017, HoEtAl2018, KuCh2017}, though interpreted slightly differently for this key feature of the exciton-plasmon interaction. We here take an opportunity to demonstrate these effects with rigorous analytical and numerical calculations, explaining how one can control them with a two-component SWCN array film.

We show that for a two-component closely packed mixture of two periodic arrays of identical semiconducting SWCNs, in which the nanotubes of one array differ in diameter just slightly from the nanotubes of another array, the exciton resonance of one SWCN array and the interband plasmon resonance of the other array have a minimal excitation energy separation, which makes the exciton-plasmon coupling possible in the composite mixture, providing an extra tool to control the absorption and/or emission properties and thus unveiling a path towards new tunable optoelectronic device applications~\cite{BoTaWo2010, BoTaWo2009, BoWoTa2009, BoAn2012, Bon2012, BoLa2004, BoETAL2014, ZhEtAL2006, BoPo2017, DiEtal2019, CaEtal2018}. By varying the constituent array fractions, one can tune the exciton-plasmon interaction strength, which is useful in many applications such as electroplasmonic switch~\cite{DiEtal2019} as well as quantum information processing, low threshold laser, light-emitting diodes, and solar cells~\cite{CaEtal2018}.

\section{\label{sec:2} Dielectric Response of Ultrathin CN Films}

\begin{figure}[t!]%
\begin{center}
\begin{center}
\includegraphics[width=0.975\linewidth]{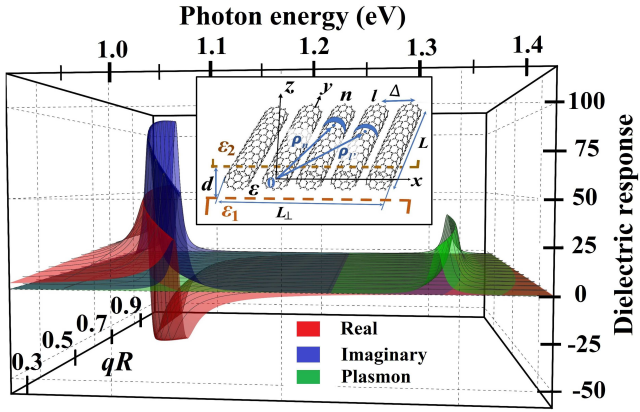}
\end{center}%
\caption{\label{fig1}%
The dielectric response functions of the (10,0) CN film of thickness $d=3R$, where $R=3.91\;$\AA~is the radius of (10,0) CN, along the CN alignment direction as functions of $qR$ and the photon energy (in eV). All graphs are calculated from Eq.~(\ref{Epyy:sigma:dl:I0Ko}) for the energies in the neighborhood of the first exciton resonance of the (10,0) CN. The fractional density $f_{_{\rm CN}}\!=\pi/9$, and $\varepsilon=10$, $\varepsilon_1=\varepsilon_2=1$. The red and blue colored surfaces depict the refraction band (${\rm Re}\,\varepsilon_\parallel$) and the exciton absorption resonance (${\rm Im}\,\varepsilon_\parallel$), respectively. The green colored surface shows the interband plasmon resonance ($-{\rm Im}\,1/\varepsilon_\parallel$). The inset shows a schematic of the geometry of relevance.}%
\end{center}
\end{figure}

Absorption of a photon of an external light source excites an electron from the valence to conduction band of a CN, thereby creating an electron-hole pair, an exciton, which induces a dipole moment polarizing the CN. An exciton in a CN can then excite the other exciton of a neighboring CN via the dipole-dipole coupling between them. One can obtain the collective polarization resulting from the dipole-dipole interaction among the CNs in a periodic CN array (Fig.~\ref{fig1}, inset) whereby the in-plane dielectric response tensor of the array can be deduced.

%           Figure 1
%

The following steps can be taken to determine the dielectric response tensor of the periodic SWCN array. The first step is to calculate the longitudinal conductivity $\sigma_{yy}(\omega)$  (the main component of the conductivity tensor) of an individual SWCN as a function of photon frequency $\omega$. This can be done by using the $(\bm{k\cdot p})$ band-structure calculation method for CNs~\cite{Ando2005}. In performing this step, as a matter of convenience, we express the result in the dimensionless form $\overline{\sigma}_{yy}(x)$ by dividing $\sigma_{yy}(\omega)$ by $e^2/(2\pi \hbar)$, where $x=\hbar\omega/(2\gamma_0)$ is the dimensionless energy in units of $2\gamma_0$ with $\gamma_0\!=\!2.7$~eV being the carbon-carbon nearest neighbor overlap integral~\cite{BoWoTa2009}. Next, we evaluate the collective polarization of the homogeneous periodic SWCN array that results from the induced dipole-dipole coupling among the individual CNs of the array. We use the low-energy plasmonic response calculation technique~\cite{Bo2019ME} combined with the many-particle Green's function formalism~\cite{Mahan2000} to relate the CN longitudinal polarizability per unit length (expressed in terms of the longitudinal conductivity~\cite{BoETAL2005}) to the collective polarization of the periodic CN array. The actual derivation is analytically involved and therefore will be presented in full detail separately~\cite{AdBo_PR2020}. The dielectric response tensor of the CN array is then determined using its standard relationship with the collective polarization tensor, and one can convince oneself that in this model the in-plane transverse component of the dielectric response (perpendicular to the CN alignment direction) is nothing but the static permittivity of the host dielectric layer, i.e., $\epsilon_\perp\!\!=\epsilon$. The longitudinal component of the dielectric response is controlled by the CN longitudinal polarizability (or conductivity) and takes the form as follows~\cite{AdBo_PR2020}
\begin{align}\label{Epyy:sigma:dl:I0Ko}
\frac{\epsilon_\parallel(q, x)}{\epsilon}=1- \frac{2\; f_{_{\rm CN}}\, \overline{\sigma}_{yy}(x)}
{f_{_{\rm CN}}\,\overline{\sigma}_{yy}(x)+{\rm i}x \, \frac{ \gamma_0 R\epsilon}{2 e^2} \;
\frac{1+\left(\epsilon_1+\epsilon_2\right)/(q \epsilon d)}{q R\; I_0(qR)\, K_0(qR)}} \,.
\end{align}
Here, the parameter $f_{_{\rm CN}}$ is a fraction indicating the relative volume occupied by the CN array in the dielectric layer of the film, i.e., $f_{_{\rm CN}}\!=\!N_\perp V_{_{CN}}/V\!=\!\pi R^2/d\Delta\,$, where $V_{_{CN}}$ is the volume of an individual CN of radius $R$, $N_\perp$ is the total number of CNs, $V$ is the volume of the film (dielectric layer embedding the array), $d$ is the film thickness, $\Delta$ is the center-to-center distance between two adjacent CNs, $I_0(qR)$ and $ K_0(qR)$ are the zeroth-order modified cylindrical Bessel functions with $q$ being the longitudinal quasimomentum to provide the correct normalization of the electron density distribution over cylindrical surfaces. The fraction $f_{_{\rm CN}}$ is smaller for a film with a sparse distribution of CNs and larger for the film with the close-packed CNs distribution. More explicitly, $f_{_{\rm CN}}$ satisfies the inequality $0<\!f_{_{\rm CN}}\!\le\!\pi/4$ and is always less than unity. Therefore, the screening effect of the dielectric background cannot be eliminated completely no matter how closely the CNs are packed and how thin the film is. Even if the CNs in the array are tightly packed, i.e, $\Delta\!=\!2R$ and the film (the dielectric layer) has the minimum possible thickness $d=2R$, the screening effect of the dielectric background is still present and increases if the thickness $d$ and/or the center-to-center intertube distance $\Delta$ increase, thereby reducing the resonance peak intensities in the dielectric response. Equation~\eqref{Epyy:sigma:dl:I0Ko} links the complex-valued axial conductivity along the axis of an individual CN, the longitudinal quasimomentum $q$, and the volume fraction of CNs to the complex-valued dielectric response of the entire CN film. The $q$ dependence of $\epsilon_\parallel(q,x)$ in Eq.~\eqref{Epyy:sigma:dl:I0Ko} makes the film dielectric response a strongly spatially dispersive non-local function, which can be controlled by adjusting the fraction $f_{_{\rm CN}}$ and the dielectric parameters of the CN film.

\begin{figure}[t!]%
\begin{center}
\begin{center}
\includegraphics[width=1.0\linewidth]{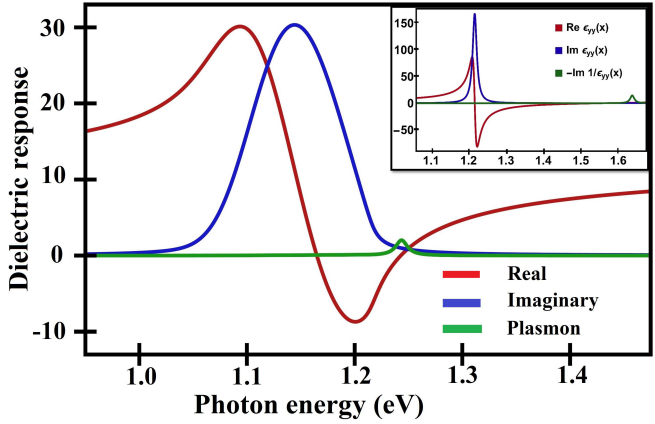}
\end{center}%
\caption{\label{Fig_Thavg}%
The thermally averaged response functions for an ultrathin film of the (10,0) CN array as functions of the photon energy. The room temperature (300 K) spectrum is calculated in the neighborhood of the first exciton resonance of the (10,0) CN for the same array parameters as those used in Fig.~\ref{fig1}. The dielectric response functions of the individual (10,0) CN are shown in the inset.}%
\end{center}
\end{figure}

One can see from Eq.~\eqref{Epyy:sigma:dl:I0Ko} that if $q\!=\!0$, then $\epsilon_\parallel(0,x)\!=\! \epsilon$. Therefore, the longitudinal component $\epsilon_\parallel (0, x)$ becomes equal to the transverse component $\epsilon_\perp(=\epsilon)$  of the CN array dielectric tensor in this limit, turning the CN film into an isotropic dielectric film. This behavior of the dielectric tensor of the CN array as $q$ approaches to zero is consistent with that of a film of metallic cylinders discussed in Ref.~\cite{Bo2019ME}. Figure~\ref{fig1} shows the dielectric responses for an ultrathin (10,0) CN film of thickness $d=3R$ and intertube distance $\Delta=3R$, calculated from Eq.~\eqref{Epyy:sigma:dl:I0Ko}. We consider the host dielectric medium of permittivity $\epsilon=10$, surrounded by air ($\epsilon_1=\epsilon_2=1$). For $d=3R$ and $\Delta=3R$ the fractional density for a homogeneous SWCN array is $f_{_{CN}}=\pi/9$, which is less than half of its the maximum possible value, indicating the presence of a significant dielectric screening effect. The red colored surface presents the refraction band defined by ${\rm Re}\,\varepsilon_\parallel$, the blue colored surface depicts the exciton absorption resonance given by ${\rm Im}\,\varepsilon_\parallel$, and the green colored surface shows the interband plasmon response defined as $-{\rm Im}\,1/\varepsilon_\parallel$, all plotted in the neighborhood of the first single-tube exciton resonance. Due to screening and spatial dispersion, in the homogeneous SWCN film the plasmon resonance is positioned much closer in energy to the exciton resonance than it occurs in an individual isolated SWCN (cf. Fig.~\ref{Fig_Thavg}, inset).

%           Figure 2

\section{\label{sec:3}Thermal Average of Dielectric Response}

To obtain the $T$-dependence of the CN array longitudinal dielectric response in Eq.~\eqref{Epyy:sigma:dl:I0Ko}, the thermal averaging can be done as follows
\begin{align}\label{Th_avg:RF}
\left<\epsilon_\parallel(T, x)\right>=\sum_q f_{\rm s} (q,T )\, \epsilon_\parallel\left(q, x\right)\,.
\end{align}
Here, the $q$-space population distribution function is of the form~\cite{AdBo_PR2020}
\begin{align}\label{Eq:fex}
f_{\rm s}(q,T)=\frac{1}{Q_{s}(T)}\;\exp \left[-\frac{\hbar\omega_s(q)}{ k_B T}\right]
\end{align}
with $\hbar\omega_s(q)=\sqrt{E_s^2(q)+2E_s(q)V_{ss}(q)}$ representing the excitation energy of the CN array collective eigen-state produced by a single-tube $s$-subband exciton with the energy $E_s(q)$ due to the intertube dipole-dipole interaction coupling $V_{ss}(q)$, and
\begin{align}\label{Eq:Qex}
Q_{s}(T)=\sum_q\exp \left[-\frac{\hbar\omega_s(q)}{ k_B T}\right]
\end{align}
to provide the proper normalization $ \sum_{q} f_{\rm s}(q,T)=1$.

Figure~\ref{Fig_Thavg} shows the thermally averaged functions ${\rm Re}\,\varepsilon_\parallel $, ${\rm Im}\,\varepsilon_\parallel $ and $-{\rm Im}\,1/\varepsilon_\parallel$ for the CN film made up of a periodic array of the (10,0) CNs. The functions are calculated numerically as given by Eqs.~(\ref{Epyy:sigma:dl:I0Ko})-(\ref{Eq:Qex}) for $T\!=300$~K and the same array parameters as those used in Fig.~\ref{fig1}. The inset in Fig.~\ref{Fig_Thavg} shows the dielectric responses of the individual (10,0) CN in the same energy range for comparison. The (10,0) CN shows sharp, narrow and intense resonance peaks, quite different from the thermally broaden resonances of the (10,0) CN array. For a single CN, the plasmon peak is positioned far away from the exciton peak. In the array of nanotubes they are spaced much closer together. The interaction among the CNs and the thermal broadening effect bring the exciton and plasmon response functions closer to each other.

%          Figure 3
%
\begin{figure}[t!]%
\begin{center}\includegraphics[width=0.975 \linewidth]{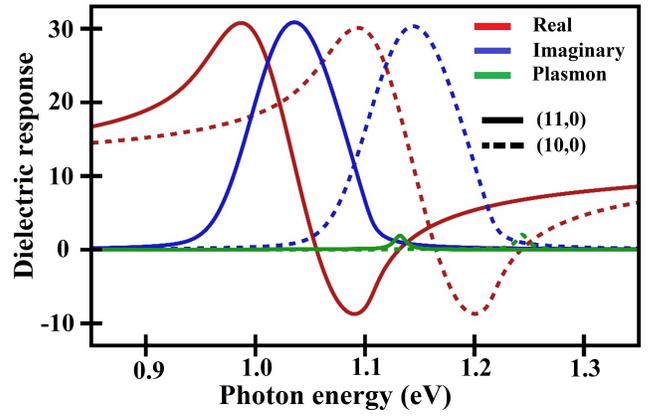}\end{center}%
\caption{\label{fig3}%
Comparison of the thermally averaged (300~K) dielectric response functions for the ultrathin arrays of the (10,0) and (11,0) CNs along the CN alignment direction. The tubules in each array are $3R$~distance apart and the films are $3R$~thick with $R$ being the respective CN radius. Other material parameters are the same as in Figs.~\ref{fig1} and~\ref{Fig_Thavg}. Only the first exciton/plasmon resonances are shown.}%
\end{figure}

Figure~\ref{fig3} compares the thermally averaged (300~K) dielectric response functions for the ultrathin (10,0) and (11,0) CN arrays, calculated from Eqs.~(\ref{Epyy:sigma:dl:I0Ko})-(\ref{Eq:Qex}) with the same array parameters as before. For both arrays the exciton and plasmon resonance peaks are situated some energy apart. However, an interesting feature here is that the plasmon peak of the (11,0) CN array is almost exactly in resonance with the exciton absorption peak of the (10,0) CN array, whereby a strong exciton-plasmon coupling should be possible in an inhomogeneous densely packed mixture of these two SWCN arrays.

To obtain the dielectric response functions for an inhomogeneous CN film composed of non-identical homogeneous SWCN arrays, the Maxwell-Garnett (MG) mixing method can be used~\cite{Ma2016}, to give the effective dielectric response $\overline{\epsilon}_\parallel(T, x)$ of the $n$-component mixture in the form
\begin{align}\label{MG:Mixing}
\overline{\epsilon}_\parallel(T, x)=\sum_{i=1}^{n}w_i\left<\epsilon_\parallel(T, x)\right>_i \,.
\end{align}
Here, $\left<\epsilon_\parallel(T, x)\right>_i$ is the thermally averaged dielectric response of the $i$-th homogeneous CN array component and $w_i$ is its relative weight within the $n$-component mixture, $\sum_{i=1}^{n}w_i\!=\!1$. The MG-mixing of the homogeneous SWCN arrays leads to an additional inhomogeneous broadening of the exciton and plasmon resonances in the CN array mixture~\cite{AdBo_MA2020}. We do believe that the broad CN film dielectric response spectra reported in Ref.~\cite{RoEtal2019} come mostly from the inhomogeneous broadening. For a properly fabricated SWCN array mixture, these broadened exciton and plasmon resonances can overlap as it occurs for the (10,0)/(11,0) CN array mixture shown in Fig.~\ref{fig3}, thereby making the exciton-plasmon coupling to strongly affect the optical properties of a composite film.

\section{\label{sec:4} Lineshape and Rabi-Splitting}

Previously, the exciton absorption lineshape under the exciton-plasmon coupling was analyzed for individual CNs and composite CN materials in Refs.~\cite{BoWoTa2009,Bo2011,NiEtAl2010,ChEtAl2017}. We here use the approach of Ref.~\cite{BoWoTa2009} for a properly tuned mixture of the two ultrathin homogeneous arrays of the (10,0) and (11,0) CNs. We mix these arrays using Eq.~\eqref{MG:Mixing}, in which we have $w_{(10,0)}+w_{(11,0)}=1$, where $w_{(10,0)}$ and $w_{(11,0)}$ are the relative weights of the thermally averaged (10,0) and (11,0) CN arrays, respectively. We focus on the energy range in the neighborhood of the first exciton absorption resonance of the (10,0) CN array. As was mentioned above and is shown in Fig.~\ref{fig3}, the interband plasmon resonance of the (11,0) CN array is almost exactly in resonance with the exciton absorption resonance of the (10,0) CN array, whereas the (10,0) CN array plasmon peak is positioned pretty much aside at higher energy and out of resonance with its own exciton absorption peak. Under these conditions, in the mixture of the two arrays the lineshape profile of the first exciton absorption resonance of the (10,0) CN array takes the form~\cite{BoWoTa2009}
\begin{equation}
I(x)=\frac{I_{0}(\epsilon_e)\,[(x-\varepsilon_e)^{2}+\Delta x_p^2]}{[(x-\varepsilon_e)^{2}-X^{2}/4]^{2}+(x-\varepsilon_e)^{2}(\Delta x_p^2+\Delta\varepsilon_e^2)}\,.
\label{Ixfin}
\end{equation}
Here, all quantities are dimensionless, i.e., normalized by $2\gamma_0$ to match Eq.~(\ref{Epyy:sigma:dl:I0Ko}) so that $\varepsilon_e\!=\!E_1^{(10,0)\!}(q)\!/2\gamma_0$, $x_p\!=\!\hbar\omega_p^{(11,0)}\!\!/2\gamma_0$, and the condition $\varepsilon_e\!\sim\!x_p$ is assumed to hold~\cite{BoWoTa2009}, which is indeed the case for the MG mixture we consider. Other notations are $I_{0}\!=\!\Gamma(\varepsilon_e)/2\pi$, $\Gamma$ is the exciton spontaneous decay rate into plasmons, $X\!=\!\sqrt{4\pi\Delta x_p\,I_{0}}$ is the Rabi-splitting parameter, $\,\Delta x_p$ is the half-width-at-half-maximum of the plasmon resonance with the energy $x_p\,$, and $\Delta\varepsilon_e$ is an additional exciton energy broadening (normally attributed to the exciton-phonon relaxation for which we use $\tau_{r\!}=\!100\;$fs, a nominal scattering time).

%           Figure 4
%
\begin{figure}[t!]%
\begin{center}\includegraphics[width=0.975 \linewidth]{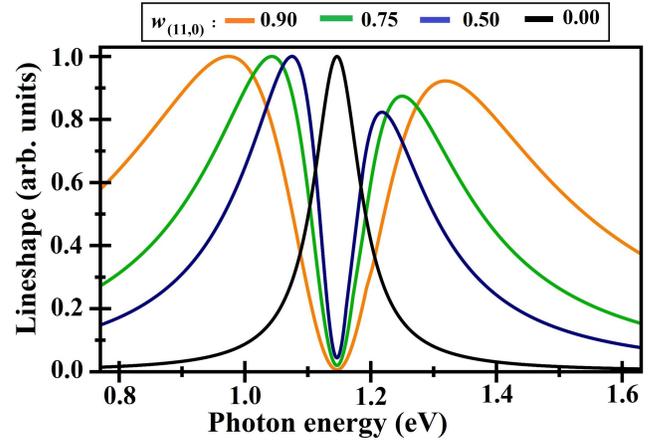}\end{center}%
\caption{\label{fig4}%
Normalized lineshape profiles of the first exciton absorption resonance of the (10,0) CN array in the MG mixture with varied relative weights of the (10,0) and (11,0) CN arrays (top). The lineshapes are first calculated using Eq.~(\ref{Ixfin}), followed by the rescaling of the two split resonance peak intensities per Eq.~(\ref{Ce2}). See text for more details.}%
\end{figure}

We mix these CN arrays at different weight combinations. From the room-temperature response functions $\mbox{Im}\,\overline{\epsilon}_\parallel(x)$ and $-\mbox{Im}\,1\!/\overline{\epsilon}_\parallel(x)$ of each mixture we determine the exact exciton and interband plasmon resonance positions, $\varepsilon_e$ and $x_p$ respectively, as well as their respective intensities and the half-width-at-half-maxima $\Delta\varepsilon_e$ and $\,\Delta x_p$ to be plugged in Eq.~(\ref{Ixfin}) to obtain the lineshape profile for the respective mixture of the (10,0) and (11,0) CN arrays forming the composite film. Furthermore, since only excitons are produced in optical dipole transitions stimulated by the transversely polarized electromagnetic (light) radiation, while plasmons require longitudinally polarized electromagnetic waves (electron beam) to be excited, the optical absorption is due to excitons only even though they might be mixed with plasmons to result in an absorption lineshape profile (Rabi) splitting. In view of this, the equal intensities of the split resonances in Eq.~(\ref{Ixfin}) should be corrected by the respective exciton and plasmon fractions, which are different from $1/2$ if the exciton $\varepsilon_e$ and plasmon $x_p$ resonance positions do not coincide exactly. More specifically, the exciton fraction is given by~\cite{Bo_OE_2015}
\begin{align}
C_e^2=\frac{1}{2}\left( 1+\frac{1\mp\sqrt{1+X^2/\delta^2}}{1+X^2/\delta^2 \mp\sqrt{1+X^2/\delta^2}}\right)\,,
\label{Ce2}
\end{align}
where $\delta=\varepsilon_e-x_p$, and the plasmon fraction is $C_p^2=1-C_e^2$, respectively. The split resonances of Eq.~(\ref{Ixfin}) should be multiplied by their respective fractions according to Eq.~(\ref{Ce2}).

Figure~\ref{fig4} shows the absorption lineshape profiles of the first exciton absorption resonance of the (10,0) CN array calculated using Eqs.~(\ref{Ixfin}) and (\ref{Ce2}) for four different relative weight combinations of the (10,0) and (11,0) CN arrays in the MG mixture of the same material and geometry parameters as before. All profiles exhibit the line-splitting (aka Rabi-splitting), which is a signature of the strong exciton-plasmon coupling~\cite{BoWoTa2009}. The lineshapes are normalized by the largest peak intensity. As the relative weight of the (11,0) CN-array increases, the split peak intensities tend to become even. The larger weight of the (11,0) CN-array also results in an increased broadening and a greater splitting of the entire absorption profile. This splitting quickly decreases with the reduction of the (11,0) CN-array relative weight, to eventually turn into a single-peak resonance absorption profile for zero relative weight of the (11,0) CN-array in the mixture.

%           Figure 5
%
\begin{figure}[t!]%
\begin{center}\includegraphics[width=0.975 \linewidth]{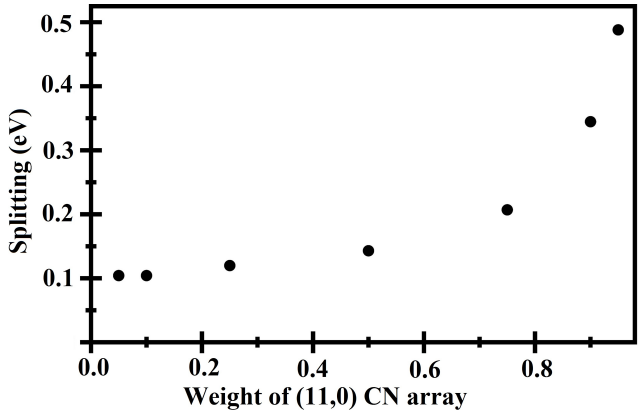}\end{center}%
\caption{\label{fig5}%
Rabi-splitting of the spectra shown in Fig.~\ref{fig4} as a function of the (11,0) CN-array component relative weight in the MG mixture.}%
\end{figure}

Figure~\ref{fig5} shows the Rabi-splitting itself of Fig.~\ref{fig4}, computed and plotted as a function of the (11,0) CN-array component relative weight. The absorption profile Rabi-splitting of the composite film can be seen to increase with the relative weight of the (11,0) CN-array, thus reflecting the increasing role of the plasmon contribution in the MG mixture. The dependence is not a linear one though; for $w_{(11,0)}>0.75$ it goes abruptly up. One can see the variation of the characteristic exciton absorption line splitting in the range of $\sim\!0.1-0.5$~eV in our case, which is as large as the typical exciton binding energies ($\sim\!0.3-0.8$~eV~\cite{Pe2004, Ped2003, Ca2006, WaEtal2005}) of individual small-diameter semiconducting CNs. The larger Rabi-splitting indicates the stronger exciton-plasmon coupling and the decreased light absorption, accordingly, in the energy window in-between the split resonance absorption peaks. Therefore, by varying the relative weights of the two array components one can tune the exciton-plasmon coupling and thereby the optical absorption profile of the composite film, to make the film transmit or absorb light in the neighborhood of an exciton absorption resonance on-demand. A very similar effect was recently demonstrated theoretically for double-wall CNs~\cite{BoPo2017} where the exciton-plasmon coupling was proposed to control the exciton Bose condensation effect in properly selected double-wall CN systems.

\section{\label{sec:5} Summary and Conclusions}

In this contribution, we study theoretically the electromagnetic response for an experimentally feasible CN film system composed of periodic arrays of parallel aligned semiconducting SWCNs embedded in an ultrathin finite-thickness dielectric layer. Our main focus is the exciton-plasmon interactions and we show how the exciton-plasmon coupling can be controlled by adjusting the intrinsic parameters of the system. We evaluate the in-plane dynamical dielectric response functions along the CN alignment direction using the low-energy plasmonic response calculation technique~\cite{Bo2019ME} combined with the many-particle Green's function formalism~\cite{Mahan2000}. The expression we obtain accounts for the intertube dipole-dipole interaction and links the response of the CN film to the complex axial conductivity of an individual constituent SWCN, an active component of the composite ultrathin film. The individual SWCN conductivity can be calculated using the $(\bm{k\cdot p})$-method of the CN-band structure theory~\cite{Ando2005}. The overall response of the system can be controlled by its intrinsic collective parameters such as SWCN volume fraction and dielectric permittivities of materials involved, in addition to the individual constituent SWCN conductivity. The CN fraction depends not only on the CN density but also on the thickness of the dielectric layer.

We also study the thermal broadening and inhomogeneity effects for the CN film dielectric response functions to understand the optical properties of realistic experimental systems at room temperature. For homogeneous single-CN films, the intertube coupling and thermal broadening bring the exciton and plasmon resonances closer together. They can even overlap due to the inhomogeneous broadening for films composed of array mixtures with a slight CN diameter distribution. In such systems the real part of the dielectric response function is negative for a sufficiently broad range of the incident photon energy (negative refraction band), and the CN film behaves as a hyperbolic metamaterial. This explains the experimental observations reported recently for horizontally aligned finite-thickness quasi-homogeneous CN films~\cite{RoEtal2019}. Using the MG mixing method~\cite{Ma2016}, we show that for a properly fabricated two-component inhomogeneous SWCN film the broadened exciton and plasmon resonances can overlap in a way it occurs for the (10,0)/(11,0) CN array mixture which we here simulate as an example. In such systems the strong exciton-plasmon coupling and associated hybridization result in the Rabi-splitting of the exciton absorption lineshape profile, thereby strongly affecting the optical response of the two-component composite film. We show that by varying the relative weights of the two array components one can tune the optical absorption profile to make the film transmit or absorb light in the neighborhood of an exciton absorption resonance on-demand.

% Figures should be put into the text as floats.

\begin{acknowledgments}
This research is supported by the U.S. National Science Foundation under Condensed Matter Theory Program Award No. DMR-1830874 (I.V.B.)\\
\end{acknowledgments}

\noindent \textbf{Data Availability:~} The data that supports the findings of this study are available within the article.


\end{document}